\documentclass[10pt,a4paper,twocolumn]{article}

\pdfoutput=1
\usepackage{ifthen}
\newboolean{uprightparticles}
\setboolean{uprightparticles}{false} 

\usepackage{amssymb}
\usepackage{amsfonts}
\usepackage{upgreek}
\usepackage{xspace}
\usepackage{color}
\usepackage{amsmath}
\usepackage{graphicx}
\usepackage{bm}
\usepackage{ctable}
\usepackage{placeins}
\usepackage{multirow}
\graphicspath{{./figs/}} 
\usepackage{hyperref}
\hypersetup{%
 colorlinks=false, linktocpage=false, pdfborder={0 0 0}, pdfstartpage=1, 
 pdfstartview=FitV,%
  }

\newboolean{articletitles}
\setboolean{articletitles}{true} 

\newboolean{pdflatex}
\setboolean{pdflatex}{true} 

\newcommand{\BRof}[1]{\ensuremath{{\cal B}(#1)}\xspace}

\newcommand{\Bs}{\ensuremath{B^0_s }\xspace}
\newcommand{\Bd}{\ensuremath{B^0}\xspace}
\newcommand{\B}{\ensuremath{B^0_{(s)} }\xspace}

\newcommand{\BuJpsimmK}{\ensuremath{B^+\to J/\psi(\mu^+\mu^-)K^+}\xspace}
\newcommand{\BuJpsieeK}{\ensuremath{B^+\to J/\psi(e^+e^-)K^+}\xspace}
\newcommand{\Jpsiee}{\ensuremath{J/\psi\to e^+ e^-}\xspace}
\newcommand{\Bhhemu}{\ensuremath{B^0_{(s)}\to h^+h^{'-}\to e^+ \mu^-}\xspace}

\def\B       {\ensuremath{B}\xspace}
\def\Bd      {\ensuremath{B^0}\xspace}
\def\Bs      {\ensuremath{B^0_s}\xspace}

\def\lhcb {LHCb\xspace}
\newcommand{\Bhh}{\ensuremath{B^0_{(s)}\to h^+{h}^{\prime -}}\xspace}
\newcommand{\IP}{\ensuremath{{\rm IP}}\xspace}

\newcommand{\BdKpi}{\ensuremath{B^0\to K^+\pi^-}\xspace}
\newcommand{\Bsemu}{\ensuremath{B^0_s\to e^{\pm}\mu^{\mp}}\xspace}
\newcommand{\Bdemu}{\ensuremath{B^0\to e^{\pm}\mu^{\mp}}\xspace}
\newcommand{\Bemu}{\ensuremath{\ensuremath{B^{0}_{(s)}}\to e^{\pm} \mu^{\mp}}\xspace}

\newcommand{\bbdil}{\ensuremath{b\bar{b}\to l^{\pm} l'^{\mp} X}\xspace}
\newcommand{\bbem}{\ensuremath{b\bar{b}\to e^{\pm} \mu^{\mp} X}\xspace}

\newcommand{\CLs}{\ensuremath{\textrm{CL}_{\textrm{s}}}\xspace}

\newcommand{\gevc}{\ensuremath{{\mathrm{\,Ge\kern -0.1em V\!/}c}}\xspace}
\newcommand{\mevc}{\ensuremath{{\mathrm{\,Me\kern -0.1em V\!/}c}}\xspace}
\newcommand{\gevcc}{\ensuremath{{\mathrm{\,Ge\kern -0.1em V\!/}c^2}}\xspace}
\newcommand{\tevcc}{\ensuremath{{\mathrm{\,Te\kern -0.1em V\!/}c^2}}\xspace}
\newcommand{\gevgevcccc}{\ensuremath{{\mathrm{\,Ge\kern -0.1em V^2\!/}c^4}}\xspace}
\newcommand{\mevcc}{\ensuremath{{\mathrm{\,Me\kern -0.1em V\!/}c^2}}\xspace}

\def\Y#1S{\ensuremath{\Upsilon{(#1S)}}\xspace}

\newcommand\TTstrut{\rule{0pt}{3.2ex}}

\newcommand\BBstrut{\rule[-1.8ex]{0pt}{0pt}}

\def\invfb   {\ensuremath{\mbox{\,fb}^{-1}}\xspace}
\newcommand{\tev}{\ensuremath{\mathrm{\,Te\kern -0.1em V}}\xspace}

\mathchardef\PLambda="7103                 
\def\L {\ensuremath{\PLambda}\xspace}
\def\Lb{\ensuremath{\L_b^0}\xspace}

\usepackage{hyperref}
\usepackage[all]{hypcap}

\usepackage{mciteplus}

\begin{document}

\onecolumn

\begin{titlepage}
\pagenumbering{roman}

\vspace*{-1.5cm}
\centerline{\large EUROPEAN ORGANIZATION FOR NUCLEAR RESEARCH (CERN)}
\vspace*{1.5cm}
\hspace*{-0.5cm}
\begin{tabular*}{\linewidth}{lc@{\extracolsep{\fill}}r}
\vspace*{-2.7cm}\mbox{\!\!\!\includegraphics[width=.14\textwidth]
{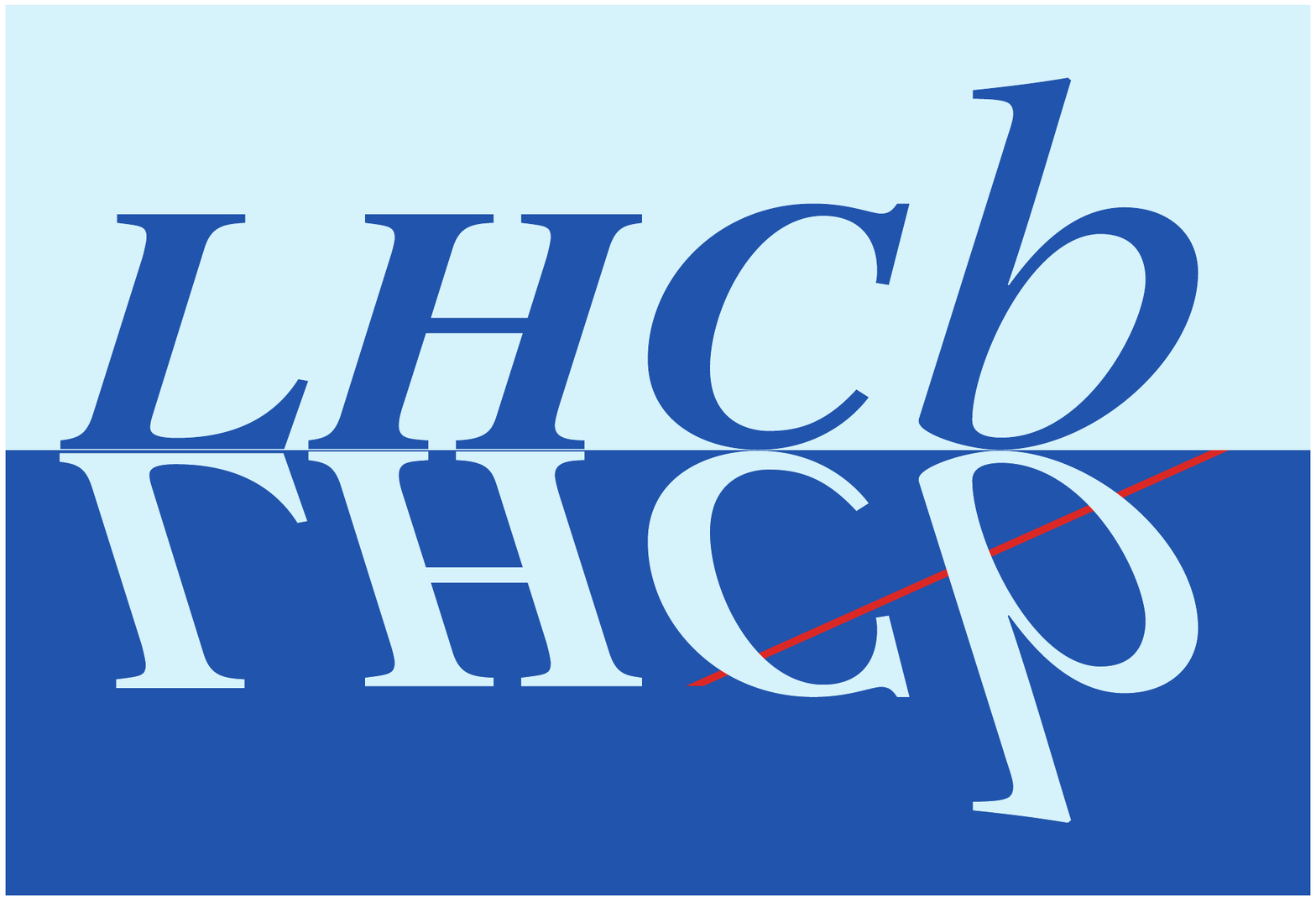}} & &
\\
 & & CERN-PH-EP-2013-123 \\  
 & & LHCb-PAPER-2013-030 \\  
 & & \\
\end{tabular*}

\vspace*{4.0cm}

{
\bf\boldmath\huge
\begin{center}
Search for the lepton-flavour violating decays 
\mbox{\boldmath $B^0_s \rightarrow e^{\pm}\mu^{\mp}$} and 
\mbox{\boldmath $B^0 \rightarrow e^{\pm} \mu^{\mp}$}
\end{center}
}

\vspace*{2.0cm}

\begin{center}
The LHCb collaboration\footnote{Authors are listed on the following pages.}
\end{center}

\vspace{\fill}

\begin{abstract}
\noindent 
A search for the lepton-flavour violating decays \Bsemu and \Bdemu is
performed with a data sample, corresponding to an integrated luminosity
of \mbox{1.0 \invfb} of $pp$ collisions at $\sqrt{s} = 7$\, TeV, collected by the LHCb experiment.
The observed number of \Bsemu and \Bdemu candidates is consistent with background expectations.
Upper limits on the branching fractions of both decays are determined to be $\BRof \Bsemu  < 1.1 \,(1.4) \times 10^{-8}$ and
$\BRof \Bdemu < 2.8 \,(3.7) \times 10^{-9}$ at 90\% (95\%) confidence level (C.L.).
These limits are a factor of twenty lower than those set by  previous experiments. Lower bounds on the Pati-Salam leptoquark masses
are also calculated, \mbox{$M_{\rm LQ} (\Bsemu) > 101$ \tevcc} and \mbox{$M_{\rm LQ} (\Bdemu) > 126$ \tevcc} at 95\% C.L.,
and are a factor of two higher than the previous bounds.
\end{abstract}

\begin{center}
  Published on Phys.~Rev.~Lett. 111, 141801 (2013) 
\end{center}

\vspace{\fill}

{\footnotesize 
\centerline{\copyright~CERN on behalf of the \lhcb collaboration, license \href{http://creativecommons.org/licenses/by/3.0/}{CC-BY-3.0}.}}
\vspace*{2mm}

\end{titlepage}



\newpage
\setcounter{page}{2}
\mbox{~}
\newpage

\centerline{\large\bf LHCb collaboration}
\begin{flushleft}
\small
R.~Aaij$^{40}$, 
B.~Adeva$^{36}$, 
M.~Adinolfi$^{45}$, 
C.~Adrover$^{6}$, 
A.~Affolder$^{51}$, 
Z.~Ajaltouni$^{5}$, 
J.~Albrecht$^{9}$, 
F.~Alessio$^{37}$, 
M.~Alexander$^{50}$, 
S.~Ali$^{40}$
% ...
\\
\bigskip
{\footnotesize \it
$ ^{1}$Centro Brasileiro de Pesquisas F\'{i}sicas (CBPF), Rio de Janeiro, Brazil\\
% ...
$ ^{5}$Clermont Universit\'{e}, Universit\'{e} Blaise Pascal, CNRS/IN2P3, LPC, Clermont-Ferrand, France\\
$ ^{6}$CPPM, Aix-Marseille Universit\'{e}, CNRS/IN2P3, Marseille, France\\
% ...
$ ^{9}$Fakult\"{a}t Physik, Technische Universit\"{a}t Dortmund, Dortmund, Germany\\
% ...
$ ^{36}$Universidad de Santiago de Compostela, Santiago de Compostela, Spain\\
$ ^{37}$European Organization for Nuclear Research (CERN), Geneva, Switzerland\\
% ...
$ ^{40}$Nikhef National Institute for Subatomic Physics, Amsterdam, The Netherlands\\
% ...
$ ^{45}$H.H. Wills Physics Laboratory, University of Bristol, Bristol, United Kingdom\\
% ...
$ ^{50}$School of Physics and Astronomy, University of Glasgow, Glasgow, United Kingdom\\
$ ^{51}$Oliver Lodge Laboratory, University of Liverpool, Liverpool, United Kingdom\\
% ...
$ ^{s}$Scuola Normale Superiore, Pisa, Italy\\
}
\end{flushleft}

\cleardoublepage
\twocolumn

\renewcommand{\thefootnote}{\arabic{footnote}}
\setcounter{footnote}{0}

\pagestyle{plain}
\setcounter{page}{1}
\pagenumbering{arabic}

\noindent
Rare decays that are forbidden in the Standard Model (SM) probe potential contributions from new processes and 
particles at a scale beyond the reach of direct searches. The decays  \mbox{\Bsemu} and \mbox{\Bdemu} and 
their charged conjugate processes\footnote{Inclusion of charge conjugate processes are implied throughout this Letter.} 
are forbidden within the SM, in which lepton flavour is conserved.
These decays are allowed in some scenarios beyond the SM that include models 
with heavy singlet Dirac neutrinos \cite{ilakovic}, 
supersymmetric models  \cite {susy} and the Pati-Salam model \cite{patisalam}. 
The latter  predicts a new interaction to mediate transitions between
leptons and quarks via exchange of spin$-$1 gauge bosons, called Pati-Salam leptoquarks (LQ), that carry both colour and lepton
quantum numbers.

Current limits from ATLAS~\cite{atlas1, atlas2, atlas3} and CMS~\cite{cms1,cms2,cms3} on the masses of first, second or 
third generation leptoquarks are in the range $[0.4, 0.9]$~\tevcc, depending on the value of the couplings and the decay channel. 
These leptoquarks arise from a coupling between a quark and lepton of the same generation.
The decays \Bsemu and \Bdemu can be mediated by other leptoquarks which couple  leptons and quarks that 
are not necessarily 
from the same generation~\cite{valencia, blanke}, such as when the $\tau$ lepton couples to a first or second quark generation.

The previous best upper limits on the branching fraction of these decays come from the CDF collaboration~\cite{bemu_cdf}, 
\mbox{\BRof \Bsemu $< 2.0 \; (2.6) \times 10^{-7}$} and \mbox{\BRof \Bdemu $<6.4 \; (7.9) \times 10^{-8}$} at 90\% (95\%) confidence level (C.L.).
These limits correspond to bounds on the masses of the corresponding Pati-Salam
leptoquarks of $M_{\rm LQ}(\Bsemu)> 47.8 \,(44.9)$ \tevcc
and $M_{\rm LQ}(\Bdemu) > 59.3 \,(56.3)$ \tevcc at 90 (95)\% C.L. \cite{bemu_cdf}.

\vskip 2mm
This Letter presents a search for the \Bsemu and \Bdemu lepton-flavour violating (LFV) decays performed 
with  a data sample, corresponding to an integrated luminosity
of \mbox{1.0\, \invfb}  of $pp$ collisions at $\sqrt{s} = 7$\, TeV, collected by the LHCb experiment in 2011 at the Large Hadron Collider.
To avoid potential bias, events in the signal mass region  $[5.1, 5.5] \,\gevcc$
were not examined until all analysis choices were finalized.

The LHCb detector is a single-arm forward spectrom-
eter covering the pseudorapidity range $2 <\eta < 5$, and is
described in detail in Ref.~\cite{LHCbdetector}.
Events were simulated  for this analysis using the software described 
in Refs.~\cite{Sjostrand:2006za, Lange:2001uf, Allison:2006ve, Agostinelli:2002hh, 
Golonka:2005pn, LHCb-PROC-2011-005, LHCb-PROC-2011-006}. 

The trigger~\cite{trigger} consists of a hardware stage (L0), based on information from the calorimeter 
and muon systems, followed by a
software stage (HLT) that applies a full event reconstruction, and is split into two stages called HLT1 and HLT2.
Candidate \Bemu decays considered in this analysis must satisfy a hardware decision that
requires the presence of a muon candidate with transverse momentum 
$p_{\rm T}> 1.5$\gevc.

All tracks considered in the HLT1 are required to have \mbox{$p_{\rm T}> 0.5$\gevc}.
The muon track of the \Bemu candidates is required to have \mbox{$p_{\rm T}>1.0$\gevc} and impact parameter, \mbox{\IP$>0.1$\,mm}.
The HLT2 consists of
exclusive, cut-based triggers for $B^0_{(s)}$ two-body decays, and inclusive multivariate~\cite{trigger,BBDT} $b$-hadron triggers. 

The \BdKpi decay is used as the normalization channel and  \Bhh ($h^{(\prime)} = K, \pi$) decays are used as a control channel,
since both have the same event topology as the signal. 
The \BdKpi yield is computed from the yield of \Bhh  decays, and the 
fraction of \BdKpi in the \Bhh sample, as described in Ref.~\cite{roadmap}. 
In order to minimize the bias introduced by the trigger requirements,
only  \Bhh candidates that are triggered independently of the presence of either of the two signal hadrons at L0 and HLT1 
are considered. 

The \Bemu candidates that pass the trigger selection criteria are further required to have well identified electron  and 
muon~\cite{muonid} candidates.
The measured momenta of the electrons are corrected to account for loss of momentum by bremsstrahlung in the detector using the photon
energy deposition in the electromagnetic calorimeter~\cite{Kstee}.
The signal candidates are required to be displaced with respect to any  $pp$ collision vertex (PV), and 
form a secondary vertex (SV) with $\chi^2$ per degree of freedom smaller than 9 and separated from the PV in the downstream direction by a flight distance significance greater than 15.

Only $B^0_{(s)}$ candidates with an impact parameter $\chi^2$ ($\chi^2_{\rm IP} $)  less than 25 are considered.
The $\chi^2_{\rm IP} $ of a $B^0_{(s)}$ candidate
is defined as the difference between the $\chi^2$
of the PV reconstructed with and without the considered candidate.
When more than one PV is reconstructed, that giving the smallest 
$\chi^2_{\rm IP} $ for the $B^0_{(s)}$ candidate is chosen. 
Only  $B^0_{(s)}$ candidates 
with invariant mass in the range $[4.9, 5.9]\gevcc$ are kept for further analysis. 
The selection criteria for the  \mbox{\Bhh} and  \mbox{\BdKpi} candidates are identical to those of the signal, apart from those used for particle identification.

A two-stage multivariate selection based on boosted decision trees (BDT)~\cite{Breiman,AdaBoost} 
is applied to the \Bemu candidates following the same strategy as Ref.~\cite{bsmumu}.
The two multivariate discriminants are trained using simulated samples,
\mbox{\Bsemu} for signal and \mbox{\bbdil} for background (where $l^{(\prime)}$ can either be a $\mu$ or an $e$ and 
$X$ is any other set of particles), 
which is dominated by simultaneous  semileptonic decays  of both $b$ and $\overline{b}$ hadrons within the same event.

The requirement on the first multivariate discriminant~\cite{bsmumu}
removes 75\,\% of the background while retaining 93\,\% of signal, 
as determined from simulation
using half of the available samples to train and the other half to evaluate the efficiencies.
The same selection is applied to the \BdKpi normalization channel 
and the efficiencies of this requirement for the signal and normalization channel are 
equal within 1.2\,\%, as determined from simulation.

The surviving background mainly comprises random combinations of electrons and muons from semileptonic
\bbem decays. 
In total 5766 electron-muon pairs pass the trigger, the offline selection and the first multivariate discriminant requirements.
The selected candidates are classified in a binned two-dimensional space formed by the electron-muon invariant mass
and the output of a second BDT, for which nine variables are employed~\cite{bsmumu}.
The BDT output is independent of the invariant mass for signal 
inside the search window. The output is transformed such that the signal is approximately uniformly 
distributed  between zero and one, while the background  peaks at zero.

The probability for a signal event to have a given BDT value is 
obtained from data using the \Bhh sample~\cite{bsmumu_plb, bsmumu_prl}.
Simulated samples of  \Bemu  and \Bhh decays have been used to check that the distributions of the variables entering in the BDT
that do not depend on the bremsstrahlung radiation are in good agreement. Corrections to the BDT shape due to the presence 
of the radiation emitted by the electron of the  \Bemu decays have been evaluated using simulation.
The number of \Bhh signal events in each BDT bin is determined by 
fitting the $h^+h'^-$ invariant mass distribution.
The systematic uncertainty on the signal BDT probability distribution function is taken to be the maximum spread 
in the fractions of yields going into each bin, obtained by fitting the same \Bhh dataset with different signal and background fit models.
Corrections are applied to the BDT shape in order to take into account the effect of the 
different trigger requirements used for the signal and the \Bhh control sample.

The invariant mass line shape of the signal events is described 
by a Crystal Ball function (CB)~\cite{crystalball} with two tails, left and right, 
defined by two parameters each. The values of the parameters 
depend on the momentum resolution, 
the momentum scale and the amount of bremsstrahlung radiation recovered.

The signal shape parameters are obtained from simulation, but need to be reweighted to account
for their dependency on the event multiplicity,
which affects the amount of bremsstrahlung radiation recovered and differs between data and simulation.  
We use the number of hits in the scintillating pad detector ($N_{\rm SPD}$) as a measure of the event multiplicity. The distribution of 
$N_{\rm SPD}$ for \Bemu  signal
candidates is obtained from a \BuJpsimmK
data sample, which is selected with the same trigger conditions as the signal,
ensuring a similar distribution of $N_{\rm SPD}$. 
The signal mass shape parameters are determined by reweighting the \Bemu simulated events with the 
$N_{\rm SPD}$ distribution measured in the \BuJpsimmK sample.

This reweighting technique is used also for a \mbox{\Jpsiee} simulated sample and the reweighted parameters are then compared
with those obtained with a \mbox{\Jpsiee} sample in  data. 
The difference between the mean values of the \mbox{\Jpsiee} mass in data and simulation (+0.16\%) is applied as a 
systematic shift to the peak values of the \Bdemu and \Bsemu invariant mass in simulation. A systematic uncertainty is added to the \Bemu 
mass parameters when the differences in the values of the other mass parameters for the \Jpsiee sample in data and SPD-reweighted simulation 
are larger than their statistical uncertainties.

The signal region, defined by the invariant mass window $[5.1,5.5]$\gevcc, retains
\mbox{$(85.0 \pm 0.1_{\rm stat} \pm 5.0_{\rm syst}) \%$} and \mbox{$(82.0 \pm 0.1_{\rm stat} \pm 5.0_{\rm syst}) \%$}
of the \Bsemu and \Bdemu signal decays, respectively. The systematic uncertainties on these fractions are evaluated
with pseudo-experiments that fluctuate each parameter of the mass lineshape according to its uncertainty. 
The width of the corresponding fraction distribution is taken as the systematic uncertainty.

The \Bsemu and \Bdemu yields are translated 
into branching fractions according to
\begin{eqnarray}
\BRof \Bemu &=&  
\frac{{\cal B}_{\rm norm}  \,{\rm \epsilon_{\rm norm}}\,f_{d} }{ N_{\rm norm}\,{\rm \epsilon_{sig}} \,f_{d(s)} } \times
N_{\Bemu}  \nonumber \\
& = & \alpha_{B^0_{(s)}} \times N_{\Bemu},
\label{eq:normalization}
\end{eqnarray}
where $N_{\rm norm} = 10\,120\pm 920$ is the number of signal events in the normalization 
channel and is determined from the total yield of the \Bhh channel and the fraction of \BdKpi events in the inclusive sample. 
The systematic uncertainty is comparable to the statistical one and 
is dominated by the maximum spread in the yield obtained by fitting the same
\Bhh dataset with different fit models \cite{bsmumu_plb, bsmumu_prl}.
The branching fraction of the normalization channel is ${\cal B}_{\rm norm}$ $= (1.94 \pm 0.06) \times 10^{-5}$~\cite{PDG2012} 
and $N_{\Bemu}$ is the number of observed signal events.
The factors $f_{d}$ and $f_{s}$ indicate the probabilities
that a $b$ quark fragments into a $B^0$ or $B^0_s$ meson, respectively.
We use $f_s/f_d = 0.256 \pm 0.020$ 
measured in $pp$ collision data at $\sqrt{s}=7$ TeV~\cite{LHCb-PAPER-2012-037}.
The measured dependence of $f_s/f_d$ on the \B meson $p_{\rm T}$~\cite{LHCb-PAPER-2012-037} 
is found to be negligible for this analysis.

The efficiency ${\rm \epsilon_{sig(norm)}}$ for the signal (normalization) channel is
the product of the reconstruction efficiency of the final state particles 
including the geometric detector acceptance, 
the selection efficiency and the trigger efficiency. 
The ratios of acceptance, reconstruction and selection efficiencies are computed with 
simulation. A systematic uncertainty is assigned to these ratios, to take into account the difference between the tracking efficiencies measured in data and predicted in simulation.
Reweighting techniques are used to correct distributions in the simulation that  do not match 
those from data, in particular for those variables that depend on  $N_{\rm SPD}$.
The trigger efficiency of L0 and HLT1  on signal decays is evaluated using data, while the HLT2 efficiency is evaluated 
in simulation after validation with control samples. 
The electron and muon identification efficiencies are evaluated from data using the \BuJpsimmK and \BuJpsieeK control samples.
The two normalization factors $\alpha_{\Bs}$ and $\alpha_{\Bd}$ are determined to be
$\alpha_{\Bs}= (1.1 \pm 0.2) \times 10^{-9}$ and
$\alpha_{\Bd}= (2.8 \pm 0.5) \times 10^{-10}$.

The BDT range is divided into eight bins with boundaries at $0.0,0.25,0.4,0.5,0.6,0.7,0.8,0.9$ and 1.0. 
The number of expected combinatorial background events in each BDT bin and the invariant mass signal region
is determined from data by fitting to an exponential function events in 
the mass sidebands, defined by $[4.9, 5.0] \gevcc$ and $[5.5,  5.9] \gevcc$.
The invariant mass distributions of the selected candidates in BDT bins and the binned BDT distributions for the signals and the combinatorial background samples are available 
in Supplemental Material~\cite{Figure_bsdemu_mass}.

In the exponential function both the slope and the normalization are allowed to vary.
The systematic uncertainty on the estimated number of combinatorial 
background events in the signal regions is determined 
by fluctuating the number of events observed in the sidebands according to a Poisson distribution, 
and by varying the exponential slope according to its uncertainty.
As a cross-check, two other models, 
the sum of two exponential functions and a single exponential fitted to the right sideband only,
have been used and provide  consistent background estimates inside the signal region. 

The low-mass sideband and the signal region are potentially polluted by exclusive backgrounds.
The background from \mbox{$B^+_c \to J/\psi(\mu^+ \mu^-) e^+ \nu_{e}$} and 
\mbox{$B^+_c \to J/\psi(e^+ e^-) \mu^+ \nu_{\mu}$} 
decays is evaluated assuming the branching fraction 
value from Ref.~\cite{Abe:1998wi}. 
The decays $B^0 \to \pi^- l^+ \nu_{l}$, \Bhh,
$B^0_s \to K^- l^+ \nu_{l}$, $\Lb \to p l^- \overline{\nu}_{l}$  and $B^{+} \to \pi^{+} l^+ l^-$  (where $l^{\pm} = e^{\pm}$ or $\mu^{\pm}$)
are potential backgrounds if the hadrons are misidentified as electrons or muons.
The $B^0 \to \pi^- l^+ \nu_{l}$ and \Bhh branching fractions are taken from Ref.~\cite{PDG2012}.
The $B^{+} \to \pi^{+} l^+ l^-$  branching fraction is taken from Ref.~\cite{Bpimumu}.
The theoretical estimates of the $\Lb \to p l^- \overline{\nu}_{l}$ and $B^0_s \to K^- l^+ \nu_{l}$ 
branching fractions are taken from Refs.~\cite{Wang} and \cite{BsKmunu}, respectively. 
We use the $\Lb$ fragmentation fraction $f_{\Lb}$  measured by LHCb~\cite{Aaij:2011jp} 
and account for its $p_{\rm T}$ dependence.

The mass and BDT distributions of these background modes are evaluated from simulated samples, using the probabilities of misidentifying kaon, pion and proton  as muon or electron as functions of momenta and transverse
momenta, which are determined from $D^{*+} \to D^0 (\to K^{-} \pi^{+}) \pi^+$ and $\L \to p \pi^- $ 
data samples. 
The yield of the \Bhhemu peaking background in each BDT bin is obtained by multiplying the \Bhh yields 
obtained by fitting the invariant mass distribution of an inclusive \Bhh sample in BDT bins [29, 30] with the 
probabilities of misidentifying kaon, pion and proton as muon or electron as functions of momenta and transverse momenta, as determined from control samples.
The mass lineshape of the \Bhhemu peaking background is obtained from a simulated 
sample of doubly-misidentified \Bhh events.
Apart from \Bhh, all background modes are
normalized relative to the \BuJpsimmK decay. We assume $f_u = f_d$ where $f_u$ is the $B^+$ fragmentation fraction.

The $\Lb \to p l^- \overline{\nu}_{l}$ and the \mbox{$B^+_c \to J/\psi(\mu^+ \mu^-) e^+ \nu_{e}$} and 
\mbox{$B^+_c \to J/\psi(e^+ e^-) \mu^+ \nu_{\mu}$} modes are the dominant exclusive modes in the range 
BDT$>0.5$, where the combinatorial background is reduced by a factor $\sim 500 $ according to simulation.
These decay modes have an invariant mass distribution
that is compatible with an exponential in the region [4.9-5.9] \gevcc, and hence are taken into account by the exponential
fit to the mass sidebands. 

In the entire BDT and mass range ($[4.9, 5.9] \gevcc$), $ 4.5 \pm 0.7$ 
doubly misidentified \Bhh decays are expected, with $(87.9 \pm 0.1)\%$ lying in the signal mass interval of $[5.1, 5.5] \gevcc.$

For each BDT bin we count the number of candidates observed in the signal region, 
and compare to the expected number of signal and background events.

The systematic uncertainties in the background and signal predictions in each bin 
are computed by varying the normalization factor, and the mass and BDT shapes 
within their Gaussian uncertainties.

The results for the \Bsemu and \Bdemu decays 
are summarized in Table~\ref{tab:data_bsdemu}. In the high BDT range, the observed number of candidates is in agreement with the 
number of expected exclusive backgrounds in the signal region.
The compatibility of the observed distribution of events  
with that expected for a given branching fraction 
hypothesis is computed with the \CLs method~\cite{Read_02}.
\begin{center}
\begin{table*}[htb]
\centering
\caption[]{Expected background (bkg) from the fit to the data sidebands, and 
expected \Bhhemu events,  compared to
the number of observed events in the mass signal region, in bins of BDT response.}
\label{tab:data_bsdemu}
{\small
\begin{tabular}{ccccccccc}
\hline\hline
 BDT bin & 0.0 -- 0.25 & 0.25 -- 0.4 & 0.4 -- 0.5 & 0.5 -- 0.6 & 0.6 -- 0.7 & 0.7 -- 0.8 & 0.8 -- 0.9 & 0.9 -- 1.0 \TTstrut\BBstrut \\
\hline
   Expected bkg (from fit) & $2222 \pm 51$ & $80.9^{+10.1}_{-9.4}$ & $20.4^{+5.0}_{-4.5}$ & $13.2^{+3.9}_{-3.6}$ & $2.1^{+2.9}_{-1.4}$ & $3.1^{+1.9}_{-1.4}$ 
& $3.1^{+1.9}_{-1.4}$ & $1.7^{+1.4}_{-1.0}$  \TTstrut\\
  Expected \Bhh bkg   & 0.67$\pm$0.12 & 0.47$\pm$0.09 & 0.40$\pm$0.08 & 0.37$\pm$0.06 & 0.45$\pm$0.08 & 0.49$\pm$0.08 & 0.57$\pm$0.09 & 0.54$\pm$0.12  \TTstrut\\
  Observed & $2332$ & $90$ & $19$ & $4$ & $3$ & $3$ & $3$ & $1$  \TTstrut\BBstrut\\\hline\hline
\end{tabular}
}
\end{table*}
 \end{center}

\begin{figure*}[!htb]
\centering
\includegraphics[width=0.45\textwidth]{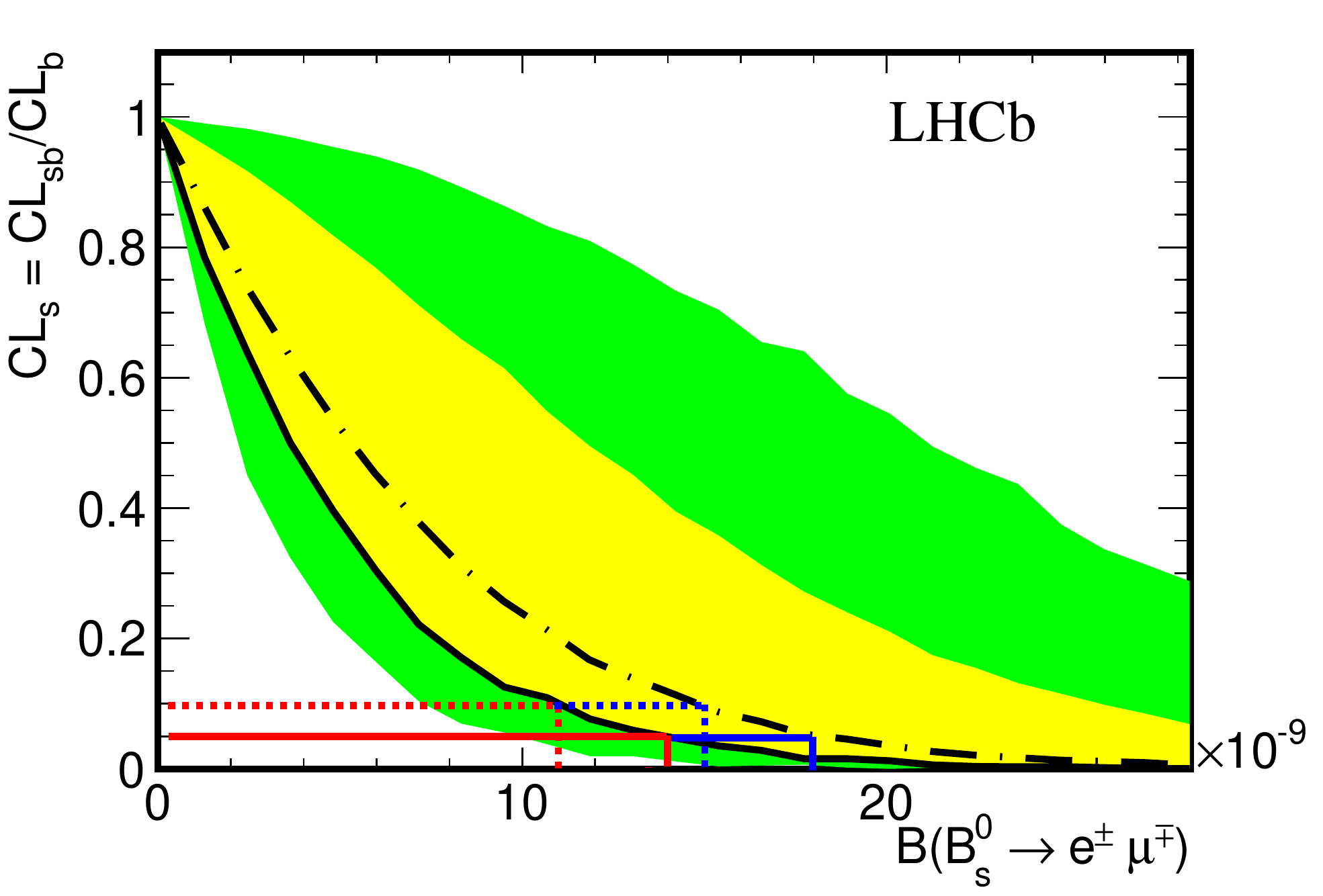}
\includegraphics[width=0.45\textwidth]{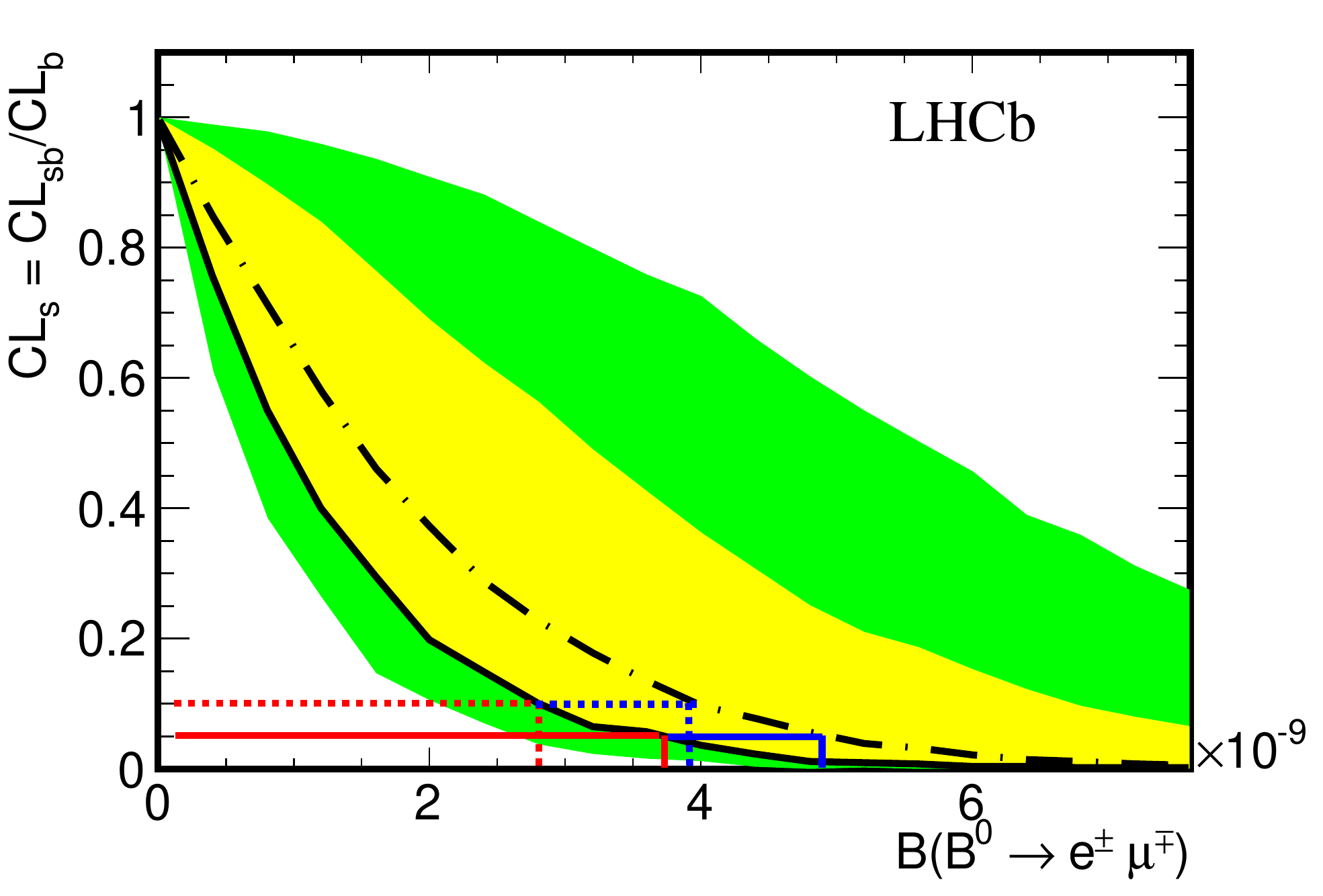}
\caption
{
 \CLs as a function of the assumed branching fraction for (left) \Bsemu and (right) 
\Bdemu decays.
The dashed lines are the medians of the expected \CLs\ distributions  if background only was observed.
The yellow (green) area covers, at a given branching fraction, 34\%(47.5\%) of the expected \CLs distribution on each side of its median.
The solid black curves are the observed \CLs. The upper limits at 90\,\% (95\,\%) C.L.  are indicated by the 
dotted (solid) vertical lines in blue for the expectation and in red for the observation.}
\label{fig:cls_bsbd}
\end{figure*}

The expected and observed \CLs values are shown in Fig.~\ref{fig:cls_bsbd} for the \Bsemu and \Bdemu channels,
as a function of the assumed branching fraction.
The expected and measured limits for \Bsemu and \Bdemu at 90\,\% 
and 95\,\% C.L. are shown in Table~\ref{tab:bds_results}.
Note that since the same events are used to set limits for both \Bs and \Bd decays, the results are strongly correlated.
The inclusion of systematic uncertainties increases the expected \Bdemu and \Bsemu upper limits by  $\sim 20\%$.
The systematic uncertainties are dominated by the uncertainty in the interpolation of the background yields 
inside the signal region.
The observed limits are $\sim 1 \, \sigma$ below the expectation due to the lower than expected numbers of observed events in the 
fourth and last BDT bins.

\begin{table}[!htb]
\caption{Expected (background only) and observed limits on the \Bemu branching fractions. 
}
\label{tab:bds_results}
\begin{center}
\begin{tabular}{llcc}
\hline\hline 
         Mode & Limit &  90\,\% C.L. &  95\,\% C.L. \TTstrut\BBstrut\\ 
\hline 
\Bsemu     & Expected                &  $1.5 \times 10^{-8} $  & $ 1.8  \times 10^{-8} $\TTstrut   \\ 
                 & Observed              &  $1.1 \times 10^{-8} $  & $ 1.4  \times 10^{-8} $   \\ 
\hline
\Bdemu      & Expected              &  $3.8 \times 10^{-9}$  & $ 4.8 \times 10^{-9}$\TTstrut \\ 
                 & Observed              &  $2.8 \times 10^{-9}$  & $ 3.7 \times 10^{-9}$   \\ 
\hline \hline
\end{tabular}
\end{center}
\end{table}

In the framework of the Pati-Salam model, the relation linking the \Bemu branching fractions and the leptoquark mass
($M_{\rm LQ}$)~\cite{valencia} is 
\begin{equation}
 \BRof \Bemu = \pi { \alpha_S^2(M_{\rm LQ}) \over M_{\rm LQ}^4} F^2_{B^0_{(s)}} m^3_{B^0_{(s)}} R^2 { \tau_{B^0_{(s)}} \over \hbar} ,
\label{eq:LQ}
\end{equation}

\noindent where 
\begin{equation}
R = { m_{B^0_{(s)}} \over m_b } {\left( \alpha_S(M_{\rm LQ}) \over \alpha_S(m_t) \right)}^{-{4 \over 7}} {\left(\alpha_S(m_t) \over \alpha_S(m_b) \right) }^{-{12\over 23}}.
\nonumber
\end{equation}

The $B^0$ and $B^0_s$ masses, $m_{B^0}$ and $m_{B^0_s}$, and the average lifetimes,  $\tau_{B^0}$ and $\tau_{B^0_s} $, are  taken from 
Ref.~\cite{PDG2012}.
The factors \mbox{$F_{B^0} = 0.190 \pm 0.004$} GeV and \mbox{$F_{B^0_s} = 0.227 \pm 0.004$} GeV 
are the decay constants of the $B^0$ and $B^0_s$ mesons~\cite{Davies}, and $m_b$ and  $m_t $ 
are the bottom and top quark masses~\cite{PDG2012}, respectively, computed in the 
$\overline{\rm MS}$ scheme~\cite{Bardeen:1978yd}.
The value of $\alpha_s$ at an arbitrary scale $M_{\rm LQ}$ is determined
using the software package  {\sc rundec}~\cite{RunDec}.

Using the limits on the branching fractions shown in Table~\ref{tab:bds_results}, we find the following lower bounds for the leptoquark masses
if the leptoquark links the $\tau$ lepton to the first and second quark generation,
\mbox{$M_{\rm LQ} (\Bsemu) > 107 \,(101) \tevcc$} and  \mbox{$M_{\rm LQ} (\Bdemu) > 135 \,(126) \tevcc$} at 90 (95)\,\% C.L., respectively.
When the parameters entering in Eq.~\ref{eq:LQ} are fluctuated within $\pm 1 \,\sigma$, the limits on the leptoquark masses change by $\sim \pm 1 $ TeV.

In summary, a search for the lepton-flavour violating decays \Bsemu
and \Bdemu has been performed on a data sample, corresponding to an
integrated luminosity of 1.0\invfb, collected in $pp$ collisions at $\sqrt{s} = 7$ TeV.
The data  are consistent with the background-only hypothesis.
Upper limits are set on the branching fractions, \mbox{\BRof \Bsemu $< 1.1 \,(1.4) \times 10^{-8}$} and 
\mbox{\BRof \Bdemu $< 2.8 \,(3.7) \times 10^{-9}$} at 90 (95)\,\% C.L., that are the most restrictive to date.
These limits translate into lower bounds on the leptoquark masses in the Pati-Salam model \cite{valencia} of  \mbox{$M_{\rm LQ} (\Bsemu) > 107 \,(101) \tevcc$} and 
\mbox{$M_{\rm LQ} (\Bdemu) > 135 \,(126) \tevcc $} at \mbox{90 (95)\,\% C.L.}, respectively. These are a factor of two higher than the previous bounds.

\section{Acknowledgements}
\noindent We thank Diego Guadagnoli for the theory inputs.
We express our gratitude to our colleagues in the CERN
accelerator departments for the excellent performance of the LHC. We
thank the technical and administrative staff at the LHCb
institutes. We acknowledge support from CERN and from the national
agencies: CAPES, CNPq, FAPERJ and FINEP (Brazil); NSFC (China);
CNRS/IN2P3 and Region Auvergne (France); BMBF, DFG, HGF and MPG
(Germany); SFI (Ireland); INFN (Italy); FOM and NWO (The Netherlands);
SCSR (Poland); MEN/IFA (Romania); MinES, Rosatom, RFBR and NRC
``Kurchatov Institute'' (Russia); MinECo, XuntaGal and GENCAT (Spain);
SNSF and SER (Switzerland); NAS Ukraine (Ukraine); STFC (United
Kingdom); NSF (USA). We also acknowledge the support received from the
ERC under FP7. The Tier1 computing centres are supported by IN2P3
(France), KIT and BMBF (Germany), INFN (Italy), NWO and SURF (The
Netherlands), PIC (Spain), GridPP (United Kingdom). We are thankful
for the computing resources put at our disposal by Yandex LLC
(Russia), as well as to the communities behind the multiple open
source software packages that we depend on.

\bibliographystyle{LHCb}
\bibliography{article_bemu}

\ifx\mcitethebibliography\mciteundefinedmacro
\PackageError{LHCb.bst}{mciteplus.sty has not been loaded}
{This bibstyle requires the use of the mciteplus package.}\fi
\providecommand{\href}[2]{#2}
\begin{mcitethebibliography}{10}
\mciteSetBstSublistMode{n}
\mciteSetBstMaxWidthForm{subitem}{\alph{mcitesubitemcount})}
\mciteSetBstSublistLabelBeginEnd{\mcitemaxwidthsubitemform\space}
{\relax}{\relax}

\bibitem{ilakovic}
A.~Ilakovic, \ifthenelse{\boolean{articletitles}}{{\it {Lepton flavor violation
  in the Standard Model extended by heavy singlet Dirac neutrinos}},
  }{}\href{http://dx.doi.org/10.1103/PhysRevD.62.036010}{Phys.\ Rev.\ D {\bf
  62} (2000) 036010}\relax
\mciteBstWouldAddEndPuncttrue
\mciteSetBstMidEndSepPunct{\mcitedefaultmidpunct}
{\mcitedefaultendpunct}{\mcitedefaultseppunct}\relax
\EndOfBibitem
\bibitem{susy}
R.~A. Diaz {\em et~al.}, \ifthenelse{\boolean{articletitles}}{{\it {Improving
  bounds on flavor changing vertices in the two Higgs doublet model from
  $B^0-\bar{B}^0$ mixing}}, }{}Eur.\ Phys.\ J {\bf C41} (2005) 305\relax
\mciteBstWouldAddEndPuncttrue
\mciteSetBstMidEndSepPunct{\mcitedefaultmidpunct}
{\mcitedefaultendpunct}{\mcitedefaultseppunct}\relax
\EndOfBibitem
\bibitem{patisalam}
J.~C. Pati and A.~Salam, \ifthenelse{\boolean{articletitles}}{{\it {Lepton
  number as the fourth color}}, }{}Phys.\ Rev.\ D {\bf 10} (1974) 275\relax
\mciteBstWouldAddEndPuncttrue
\mciteSetBstMidEndSepPunct{\mcitedefaultmidpunct}
{\mcitedefaultendpunct}{\mcitedefaultseppunct}\relax
\EndOfBibitem
\bibitem{atlas1}
ATLAS collaboration, G.~Aad {\em et~al.},
  \ifthenelse{\boolean{articletitles}}{{\it {Search for second generation
  scalar leptoquarks in $pp$ collisions at $\sqrt{s} = 7$~TeV with the ATLAS
  detector}}, }{}\href{http://dx.doi.org/10.1140/epjc/s10052-012-2151-6}{Eur.\
  Phys.\ J.\  {\bf C72} (2012) 2151},
  \href{http://arxiv.org/abs/1203.3172}{{\tt arXiv:1203.3172}}\relax
\mciteBstWouldAddEndPuncttrue
\mciteSetBstMidEndSepPunct{\mcitedefaultmidpunct}
{\mcitedefaultendpunct}{\mcitedefaultseppunct}\relax
\EndOfBibitem
\bibitem{atlas2}
ATLAS collaboration, G.~Aad {\em et~al.},
  \ifthenelse{\boolean{articletitles}}{{\it {Search for first generation scalar
  leptoquarks in $pp$ collisions at $\sqrt{s} = 7$~\mbox{TeV} with the ATLAS
  detector}}, }{}\href{http://dx.doi.org/10.1016/j.physletb.2012.02.004}{Phys.\
  Lett.\  {\bf B709} (2012) 158}, \href{http://arxiv.org/abs/1112.4828}{{\tt
  arXiv:1112.4828}}\relax
\mciteBstWouldAddEndPuncttrue
\mciteSetBstMidEndSepPunct{\mcitedefaultmidpunct}
{\mcitedefaultendpunct}{\mcitedefaultseppunct}\relax
\EndOfBibitem
\bibitem{atlas3}
ATLAS collaboration, G.~Aad {\em et~al.},
  \ifthenelse{\boolean{articletitles}}{{\it {Search for third generation scalar
  leptoquarks in pp collisions at sqrt(s) = 7 TeV with the ATLAS detector}},
  }{}\href{http://dx.doi.org/10.1007/JHEP06(2013)033}{JHEP {\bf 06} (2013)
  033}, \href{http://arxiv.org/abs/1303.0526}{{\tt arXiv:1303.0526}}\relax
\mciteBstWouldAddEndPuncttrue
\mciteSetBstMidEndSepPunct{\mcitedefaultmidpunct}
{\mcitedefaultendpunct}{\mcitedefaultseppunct}\relax
\EndOfBibitem
\bibitem{cms1}
CMS collaboration, S.~Chatrchyan {\em et~al.},
  \ifthenelse{\boolean{articletitles}}{{\it {Search for pair production of
  first- and second-generation scalar leptoquarks in $pp$ collisions at
  $\sqrt{s}= 7$~TeV}},
  }{}\href{http://dx.doi.org/10.1103/PhysRevD.86.052013}{Phys.\ Rev.\ D {\bf
  86} (2012) 052013}, \href{http://arxiv.org/abs/1207.5406}{{\tt
  arXiv:1207.5406}}\relax
\mciteBstWouldAddEndPuncttrue
\mciteSetBstMidEndSepPunct{\mcitedefaultmidpunct}
{\mcitedefaultendpunct}{\mcitedefaultseppunct}\relax
\EndOfBibitem
\bibitem{cms2}
CMS collaboration, S.~Chatrchyan {\em et~al.},
  \ifthenelse{\boolean{articletitles}}{{\it {Search for third-generation
  leptoquarks and scalar bottom quarks in $pp$ collisions at $\sqrt{s}=
  7$~TeV}}, }{}\href{http://dx.doi.org/10.1007/JHEP12(2012)055}{JHEP {\bf 12}
  (2012) 055}, \href{http://arxiv.org/abs/1210.5627}{{\tt
  arXiv:1210.5627}}\relax
\mciteBstWouldAddEndPuncttrue
\mciteSetBstMidEndSepPunct{\mcitedefaultmidpunct}
{\mcitedefaultendpunct}{\mcitedefaultseppunct}\relax
\EndOfBibitem
\bibitem{cms3}
CMS collaboration, S.~Chatrchyan {\em et~al.},
  \ifthenelse{\boolean{articletitles}}{{\it {Search for pair production of
  third-generation leptoquarks and top squarks in $pp$ collisions at
  $\sqrt{s}=7$ TeV}},
  }{}\href{http://dx.doi.org/10.1103/PhysRevLett.110.081801}{Phys.\ Rev.\
  Lett.\  {\bf 110} (2013) 081801}, \href{http://arxiv.org/abs/1210.5629}{{\tt
  arXiv:1210.5629}}\relax
\mciteBstWouldAddEndPuncttrue
\mciteSetBstMidEndSepPunct{\mcitedefaultmidpunct}
{\mcitedefaultendpunct}{\mcitedefaultseppunct}\relax
\EndOfBibitem
\bibitem{valencia}
G.~Valencia and S.~Willenbrock, \ifthenelse{\boolean{articletitles}}{{\it
  {Quark-lepton unification and rare meson decays}},
  }{}\href{http://dx.doi.org/10.1103/PhysRevD.50.6843}{Phys.\ Rev.\ D {\bf 50}
  (1994) 6843}\relax
\mciteBstWouldAddEndPuncttrue
\mciteSetBstMidEndSepPunct{\mcitedefaultmidpunct}
{\mcitedefaultendpunct}{\mcitedefaultseppunct}\relax
\EndOfBibitem
\bibitem{blanke}
M.~Blanke {\em et~al.}, \ifthenelse{\boolean{articletitles}}{{\it {Charged
  lepton flavour violation and $(g-2)_{\mu}$ in the \mbox{Littlest Higgs Model}
  with T-Parity: a clear distinction from Supersymmetry}},
  }{}\href{http://dx.doi.org/10.1088/1126-6708/2007/05/013}{JHEP {\bf 05}
  (2007) 013}, \href{http://arxiv.org/abs/0702136}{{\tt arXiv:0702136}}\relax
\mciteBstWouldAddEndPuncttrue
\mciteSetBstMidEndSepPunct{\mcitedefaultmidpunct}
{\mcitedefaultendpunct}{\mcitedefaultseppunct}\relax
\EndOfBibitem
\bibitem{bemu_cdf}
CDF collaboration, T.~Aaltonen {\em et~al.},
  \ifthenelse{\boolean{articletitles}}{{\it {Search for the decays
  ${B^0_{(s)}\to e^+ \mu^-}$ and ${B^0_{(s)}\to e^+e^-}$ in CDF Run II}},
  }{}\href{http://dx.doi.org/10.1103/PhysRevLett.102.201801}{Phys.\ Rev.\
  Lett.\  {\bf 102} (2009) 201801}, \href{http://arxiv.org/abs/0901.3803}{{\tt
  arXiv:0901.3803}}\relax
\mciteBstWouldAddEndPuncttrue
\mciteSetBstMidEndSepPunct{\mcitedefaultmidpunct}
{\mcitedefaultendpunct}{\mcitedefaultseppunct}\relax
\EndOfBibitem
\bibitem{LHCbdetector}
LHCb collaboration, A.~A. Alves~Jr. {\em et~al.},
  \ifthenelse{\boolean{articletitles}}{{\it {The \lhcb detector at the LHC}},
  }{}\href{http://dx.doi.org/10.1088/1748-0221/3/08/S08005}{JINST {\bf 3}
  (2008) S08005}\relax
\mciteBstWouldAddEndPuncttrue
\mciteSetBstMidEndSepPunct{\mcitedefaultmidpunct}
{\mcitedefaultendpunct}{\mcitedefaultseppunct}\relax
\EndOfBibitem
\bibitem{Sjostrand:2006za}
T.~Sj\"{o}strand, S.~Mrenna, and P.~Skands,
  \ifthenelse{\boolean{articletitles}}{{\it {PYTHIA 6.4 physics and manual}},
  }{}\href{http://dx.doi.org/10.1088/1126-6708/2006/05/026}{JHEP {\bf 05}
  (2006) 026}, \href{http://arxiv.org/abs/hep-ph/0603175}{{\tt
  arXiv:hep-ph/0603175}}\relax
\mciteBstWouldAddEndPuncttrue
\mciteSetBstMidEndSepPunct{\mcitedefaultmidpunct}
{\mcitedefaultendpunct}{\mcitedefaultseppunct}\relax
\EndOfBibitem
\bibitem{Lange:2001uf}
D.~J. Lange, \ifthenelse{\boolean{articletitles}}{{\it {The EvtGen particle
  decay simulation package}},
  }{}\href{http://dx.doi.org/10.1016/S0168-9002(01)00089-4}{Nucl.\ Instrum.\
  Meth.\  {\bf A462} (2001) 152}\relax
\mciteBstWouldAddEndPuncttrue
\mciteSetBstMidEndSepPunct{\mcitedefaultmidpunct}
{\mcitedefaultendpunct}{\mcitedefaultseppunct}\relax
\EndOfBibitem
\bibitem{Allison:2006ve}
GEANT4 collaboration, J.~Allison {\em et~al.},
  \ifthenelse{\boolean{articletitles}}{{\it {GEANT4 developments and
  applications}}, }{}\href{http://dx.doi.org/10.1109/TNS.2006.869826}{IEEE
  Trans.\ Nucl.\ Sci.\  {\bf 53} (2006) 270}\relax
\mciteBstWouldAddEndPuncttrue
\mciteSetBstMidEndSepPunct{\mcitedefaultmidpunct}
{\mcitedefaultendpunct}{\mcitedefaultseppunct}\relax
\EndOfBibitem
\bibitem{Agostinelli:2002hh}
GEANT4 collaboration, S.~Agostinelli {\em et~al.},
  \ifthenelse{\boolean{articletitles}}{{\it {GEANT4: a simulation toolkit}},
  }{}\href{http://dx.doi.org/10.1016/S0168-9002(03)01368-8}{Nucl.\ Instrum.\
  Meth.\  {\bf A506} (2003) 250}\relax
\mciteBstWouldAddEndPuncttrue
\mciteSetBstMidEndSepPunct{\mcitedefaultmidpunct}
{\mcitedefaultendpunct}{\mcitedefaultseppunct}\relax
\EndOfBibitem
\bibitem{Golonka:2005pn}
P.~Golonka and Z.~Was, \ifthenelse{\boolean{articletitles}}{{\it {PHOTOS Monte
  Carlo: a precision tool for QED corrections in $Z$ and $W$ decays}},
  }{}\href{http://dx.doi.org/10.1140/epjc/s2005-02396-4}{Eur.\ Phys.\ J.\  {\bf
  C45} (2006) 97}, \href{http://arxiv.org/abs/hep-ph/0506026}{{\tt
  arXiv:hep-ph/0506026}}\relax
\mciteBstWouldAddEndPuncttrue
\mciteSetBstMidEndSepPunct{\mcitedefaultmidpunct}
{\mcitedefaultendpunct}{\mcitedefaultseppunct}\relax
\EndOfBibitem
\bibitem{LHCb-PROC-2011-005}
I.~Belyaev {\em et~al.}, \ifthenelse{\boolean{articletitles}}{{\it {Handling of
  the generation of primary events in \gauss, the \lhcb simulation framework}},
  }{}\href{http://dx.doi.org/10.1109/NSSMIC.2010.5873949}{Nuclear Science
  Symposium Conference Record (NSS/MIC) {\bf IEEE} (2010) 1155}\relax
\mciteBstWouldAddEndPuncttrue
\mciteSetBstMidEndSepPunct{\mcitedefaultmidpunct}
{\mcitedefaultendpunct}{\mcitedefaultseppunct}\relax
\EndOfBibitem
\bibitem{LHCb-PROC-2011-006}
M.~Clemencic {\em et~al.}, \ifthenelse{\boolean{articletitles}}{{\it {The \lhcb
  simulation application, \gauss: design, evolution and experience}},
  }{}\href{http://dx.doi.org/10.1088/1742-6596/331/3/032023}{{J.\ of Phys:
  Conf.\ Ser.\ } {\bf 331} (2011) 032023}\relax
\mciteBstWouldAddEndPuncttrue
\mciteSetBstMidEndSepPunct{\mcitedefaultmidpunct}
{\mcitedefaultendpunct}{\mcitedefaultseppunct}\relax
\EndOfBibitem
\bibitem{trigger}
R.~Aaij {\em et~al.}, \ifthenelse{\boolean{articletitles}}{{\it {The LHCb
  trigger and its performance in 2011}},
  }{}\href{http://dx.doi.org/10.1088/1748-0221/8/04/P04022}{JINST {\bf 8}
  (2013) P04022}, \href{http://arxiv.org/abs/1211.3055}{{\tt
  arXiv:1211.3055}}\relax
\mciteBstWouldAddEndPuncttrue
\mciteSetBstMidEndSepPunct{\mcitedefaultmidpunct}
{\mcitedefaultendpunct}{\mcitedefaultseppunct}\relax
\EndOfBibitem
\bibitem{BBDT}
V.~V. Gligorov and M.~Williams, \ifthenelse{\boolean{articletitles}}{{\it
  {Efficient, reliable and fast high-level triggering using a bonsai boosted
  decision tree}},
  }{}\href{http://dx.doi.org/10.1088/1748-0221/8/02/P02013}{JINST {\bf 8}
  (2013) P02013}, \href{http://arxiv.org/abs/1210.6861}{{\tt
  arXiv:1210.6861}}\relax
\mciteBstWouldAddEndPuncttrue
\mciteSetBstMidEndSepPunct{\mcitedefaultmidpunct}
{\mcitedefaultendpunct}{\mcitedefaultseppunct}\relax
\EndOfBibitem
\bibitem{roadmap}
LHCb collaboration, B.~Adeva {\em et~al.},
  \ifthenelse{\boolean{articletitles}}{{\it {Roadmap for selected key
  measurements of LHCb}}, }{}\href{http://arxiv.org/abs/0912.4179}{{\tt
  arXiv:0912.4179}}\relax
\mciteBstWouldAddEndPuncttrue
\mciteSetBstMidEndSepPunct{\mcitedefaultmidpunct}
{\mcitedefaultendpunct}{\mcitedefaultseppunct}\relax
\EndOfBibitem
\bibitem{muonid}
F.~Archilli {\em et~al.}, \ifthenelse{\boolean{articletitles}}{{\it
  {Performance of the Muon Identification at LHCb}},
  }{}\href{http://arxiv.org/abs/1306.0249}{{\tt arXiv:1306.0249}}, submitted to
  JINST\relax
\mciteBstWouldAddEndPuncttrue
\mciteSetBstMidEndSepPunct{\mcitedefaultmidpunct}
{\mcitedefaultendpunct}{\mcitedefaultseppunct}\relax
\EndOfBibitem
\bibitem{Kstee}
LHCb collaboration, R.~Aaij {\em et~al.},
  \ifthenelse{\boolean{articletitles}}{{\it {Measurement of the $B^0
  \rightarrow K^{*0}e^+e^-$ branching fraction at low dilepton mass}},
  }{}\href{http://dx.doi.org/10.1007/JHEP05(2013)159}{JHEP {\bf 05} (2013)
  159}, \href{http://arxiv.org/abs/1304.3035}{{\tt arXiv:1304.3035}}\relax
\mciteBstWouldAddEndPuncttrue
\mciteSetBstMidEndSepPunct{\mcitedefaultmidpunct}
{\mcitedefaultendpunct}{\mcitedefaultseppunct}\relax
\EndOfBibitem
\bibitem{Breiman}
L.~Breiman, J.~H. Friedman, R.~A. Olshen, and C.~J. Stone, {\em Classification
  and regression trees}, Wadsworth international group, Belmont, California,
  USA, 1984\relax
\mciteBstWouldAddEndPuncttrue
\mciteSetBstMidEndSepPunct{\mcitedefaultmidpunct}
{\mcitedefaultendpunct}{\mcitedefaultseppunct}\relax
\EndOfBibitem
\bibitem{AdaBoost}
R.~E. Schapire and Y.~Freund, \ifthenelse{\boolean{articletitles}}{{\it A
  decision-theoretic generalization of on-line learning and an application to
  boosting}, }{}\href{http://dx.doi.org/10.1006/jcss.1997.1504}{Jour.\ Comp.\
  and Syst.\ Sc.\  {\bf 55} (1997) 119}\relax
\mciteBstWouldAddEndPuncttrue
\mciteSetBstMidEndSepPunct{\mcitedefaultmidpunct}
{\mcitedefaultendpunct}{\mcitedefaultseppunct}\relax
\EndOfBibitem
\bibitem{bsmumu}
LHCb collaboration, R.~Aaij {\em et~al.},
  \ifthenelse{\boolean{articletitles}}{{\it {First evidence for the decay
  $B^0_s \to \mu^+ \mu^-$}},
  }{}\href{http://dx.doi.org/10.1103/PhysRevLett.110.021801}{Phys.\ Rev.\
  Lett.\  {\bf 110} (2012) 021801}, \href{http://arxiv.org/abs/1211.2674}{{\tt
  arXiv:1211.2674}}\relax
\mciteBstWouldAddEndPuncttrue
\mciteSetBstMidEndSepPunct{\mcitedefaultmidpunct}
{\mcitedefaultendpunct}{\mcitedefaultseppunct}\relax
\EndOfBibitem
\bibitem{bsmumu_plb}
LHCb collaboration, R.~Aaij {\em et~al.},
  \ifthenelse{\boolean{articletitles}}{{\it {Search for the rare decays $B^0_s
  \to \mu^+ \mu^-$ and $B^0 \to \mu^+ \mu^-$}},
  }{}\href{http://dx.doi.org/10.1016/j.physletb.2012.01.38}{Phys.\ Lett.\  {\bf
  B708} (2012) 55}, \href{http://arxiv.org/abs/1112.1600}{{\tt
  arXiv:1112.1600}}\relax
\mciteBstWouldAddEndPuncttrue
\mciteSetBstMidEndSepPunct{\mcitedefaultmidpunct}
{\mcitedefaultendpunct}{\mcitedefaultseppunct}\relax
\EndOfBibitem
\bibitem{bsmumu_prl}
LHCb collaboration, R.~Aaij {\em et~al.},
  \ifthenelse{\boolean{articletitles}}{{\it {Strong constraints on the rare
  decays $B_s \to \mu^+ \mu^-$ and $B^0 \to \mu^+ \mu^-$}},
  }{}\href{http://dx.doi.org/10.1103/PhysRevLett.108.231801}{Phys.\ Rev.\
  Lett.\  {\bf 108} (2012) 231801}, \href{http://arxiv.org/abs/1203.4493}{{\tt
  arXiv:1203.4493}}\relax
\mciteBstWouldAddEndPuncttrue
\mciteSetBstMidEndSepPunct{\mcitedefaultmidpunct}
{\mcitedefaultendpunct}{\mcitedefaultseppunct}\relax
\EndOfBibitem
\bibitem{crystalball}
T.~Skwarnicki, {\em {A study of the radiative cascade transitions between the
  Upsilon-prime and Upsilon resonances}}, PhD thesis, Institute of Nuclear
  Physics, Krakow, 1986,
  {\href{http://inspirehep.net/record/230779/files/230779.pdf}{DESY-F31-86-02}%
}\relax
\mciteBstWouldAddEndPuncttrue
\mciteSetBstMidEndSepPunct{\mcitedefaultmidpunct}
{\mcitedefaultendpunct}{\mcitedefaultseppunct}\relax
\EndOfBibitem
\bibitem{PDG2012}
Particle Data Group, J.~Beringer {\em et~al.},
  \ifthenelse{\boolean{articletitles}}{{\it {\href{http://pdg.lbl.gov/}{Review
  of particle physics}}},
  }{}\href{http://dx.doi.org/10.1103/PhysRevD.86.010001}{Phys.\ Rev.\  {\bf
  D86} (2012) 010001}\relax
\mciteBstWouldAddEndPuncttrue
\mciteSetBstMidEndSepPunct{\mcitedefaultmidpunct}
{\mcitedefaultendpunct}{\mcitedefaultseppunct}\relax
\EndOfBibitem
\bibitem{LHCb-PAPER-2012-037}
LHCb collaboration, R.~Aaij {\em et~al.},
  \ifthenelse{\boolean{articletitles}}{{\it {Measurement of the ratio of
  fragmentation fractions $f_s/f_d$ and dependence on $B$ meson kinematics}},
  }{}JHEP {\bf 04} (2013) 001, \href{http://arxiv.org/abs/1301.5286}{{\tt
  arXiv:1301.5286}}\relax
\mciteBstWouldAddEndPuncttrue
\mciteSetBstMidEndSepPunct{\mcitedefaultmidpunct}
{\mcitedefaultendpunct}{\mcitedefaultseppunct}\relax
\EndOfBibitem
\bibitem{Figure_bsdemu_mass}
See Supplemental Material at [URL will be inserted by publisher] for the
  invariant mass distributions of selected candidates in BDT bins and the
  binned BDT distributions for the signals and the combinatorial background
  samples.\relax
\mciteBstWouldAddEndPunctfalse
\mciteSetBstMidEndSepPunct{\mcitedefaultmidpunct}
{}{\mcitedefaultseppunct}\relax
\EndOfBibitem
\bibitem{Abe:1998wi}
CDF collaboration, F.~Abe {\em et~al.},
  \ifthenelse{\boolean{articletitles}}{{\it {Observation of the $B_c$ meson in
  $p\bar{p}$ collisions at $\sqrt{s} = 1.8$ TeV}},
  }{}\href{http://dx.doi.org/10.1103/PhysRevLett.81.2432}{Phys.\ Rev.\ Lett.\
  {\bf 81} (1998) 2432}, \href{http://arxiv.org/abs/hep-ex/9805034}{{\tt
  arXiv:hep-ex/9805034}}\relax
\mciteBstWouldAddEndPuncttrue
\mciteSetBstMidEndSepPunct{\mcitedefaultmidpunct}
{\mcitedefaultendpunct}{\mcitedefaultseppunct}\relax
\EndOfBibitem
\bibitem{Bpimumu}
LHCb collaboration, R.~Aaij {\em et~al.},
  \ifthenelse{\boolean{articletitles}}{{\it {First observation of the decay
  $B^+ \to \pi^+ \mu^+ \mu^-$}},
  }{}\href{http://dx.doi.org/10.1007/JHEP12(2012)125}{JHEP {\bf 12} (2012)
  125}, \href{http://arxiv.org/abs/1210.2645}{{\tt arXiv:1210.2645}}\relax
\mciteBstWouldAddEndPuncttrue
\mciteSetBstMidEndSepPunct{\mcitedefaultmidpunct}
{\mcitedefaultendpunct}{\mcitedefaultseppunct}\relax
\EndOfBibitem
\bibitem{Wang}
A.~Khodjamirian, C.~Klein, T.~Mannel, and Y.-M. Wang,
  \ifthenelse{\boolean{articletitles}}{{\it {Form factors and strong couplings
  of heavy baryons from QCD light-cone sum rules}},
  }{}\href{http://dx.doi.org/10.1007/JHEP09(2011)106}{JHEP {\bf 09} (2011)
  106}, \href{http://arxiv.org/abs/1108.2971}{{\tt arXiv:1108.2971}}\relax
\mciteBstWouldAddEndPuncttrue
\mciteSetBstMidEndSepPunct{\mcitedefaultmidpunct}
{\mcitedefaultendpunct}{\mcitedefaultseppunct}\relax
\EndOfBibitem
\bibitem{BsKmunu}
W.-F. Wang and Z.-J. Xiao, \ifthenelse{\boolean{articletitles}}{{\it
  {Semileptonic decays $B/B_s \to (\pi, K)(\ell^+\ell^-,\ell\nu,\nu\bar{\nu})$
  in the perturbative QCD approach beyond the leading order}},
  }{}\href{http://dx.doi.org/10.1103/PhysRevD.86.114025}{Phys.\ Rev.\  {\bf
  D86} (2012) 114025}, \href{http://arxiv.org/abs/1207.0265}{{\tt
  arXiv:1207.0265}}\relax
\mciteBstWouldAddEndPuncttrue
\mciteSetBstMidEndSepPunct{\mcitedefaultmidpunct}
{\mcitedefaultendpunct}{\mcitedefaultseppunct}\relax
\EndOfBibitem
\bibitem{Aaij:2011jp}
LHCb collaboration, R.~Aaij {\em et~al.},
  \ifthenelse{\boolean{articletitles}}{{\it {Measurement of $b$-hadron
  production fractions in 7 TeV pp collisions}}, }{}Phys.\ Rev.\  {\bf D85}
  (2012) 032008, \href{http://arxiv.org/abs/1111.2357}{{\tt
  arXiv:1111.2357}}\relax
\mciteBstWouldAddEndPuncttrue
\mciteSetBstMidEndSepPunct{\mcitedefaultmidpunct}
{\mcitedefaultendpunct}{\mcitedefaultseppunct}\relax
\EndOfBibitem
\bibitem{Read_02}
A.~L. Read, \ifthenelse{\boolean{articletitles}}{{\it {Presentation of search
  results: the CL$_{\rm s}$ technique}},
  }{}\href{http://dx.doi.org/10.1088/0954-3899/28/10/313}{J.\ Phys.\  {\bf G28}
  (2002) 2693}\relax
\mciteBstWouldAddEndPuncttrue
\mciteSetBstMidEndSepPunct{\mcitedefaultmidpunct}
{\mcitedefaultendpunct}{\mcitedefaultseppunct}\relax
\EndOfBibitem
\bibitem{Davies}
C.~Davies, \ifthenelse{\boolean{articletitles}}{{\it {Standard Model heavy
  flavor physics on the lattice}}, }{}PoS {\bf LATTICE2011} (2011) 019,
  \href{http://arxiv.org/abs/1203.3862}{{\tt arXiv:1203.3862}}\relax
\mciteBstWouldAddEndPuncttrue
\mciteSetBstMidEndSepPunct{\mcitedefaultmidpunct}
{\mcitedefaultendpunct}{\mcitedefaultseppunct}\relax
\EndOfBibitem
\bibitem{Bardeen:1978yd}
W.~A. Bardeen, A.~Buras, D.~Duke, and T.~Muta,
  \ifthenelse{\boolean{articletitles}}{{\it {Deep inelastic scattering beyond
  the leading order in asymptotically free gauge theories}},
  }{}\href{http://dx.doi.org/10.1103/PhysRevD.18.3998}{Phys.\ Rev.\  {\bf D18}
  (1978) 3998}\relax
\mciteBstWouldAddEndPuncttrue
\mciteSetBstMidEndSepPunct{\mcitedefaultmidpunct}
{\mcitedefaultendpunct}{\mcitedefaultseppunct}\relax
\EndOfBibitem
\bibitem{RunDec}
K.~Chetyrkin, J.~Kuhn, and M.~Steinhauser,
  \ifthenelse{\boolean{articletitles}}{{\it {RunDec: a Mathematica package for
  running and decoupling of the strong coupling and quark masses }},
  }{}\href{http://dx.doi.org/10.1016/S0010-4655(00)00155-7}{Comput.\ Phys.\
  Comm.\  {\bf 133} (2000) 43}, \href{http://arxiv.org/abs/hep-ph/0004189}{{\tt
  arXiv:hep-ph/0004189}}\relax
\mciteBstWouldAddEndPuncttrue
\mciteSetBstMidEndSepPunct{\mcitedefaultmidpunct}
{\mcitedefaultendpunct}{\mcitedefaultseppunct}\relax
\EndOfBibitem
\end{mcitethebibliography}

\end{document}